\documentclass[twocolumn]{jpsj3} 

\usepackage{graphicx}
\usepackage{dcolumn}
\usepackage{bm}
\usepackage{comment}
\renewcommand{\Vec}[1]{\mbox{\boldmath$#1$}}

\title{Robust Spin Fluctuations and $s\pm$ Pairing 
in the Heavily Electron Doped Iron-Based Superconductors}
\author{Katsuhiro Suzuki$^1$, Hidetomo Usui$^2$, Kazuhiko Kuroki$^2$, 
Soshi Iimura$^3$,\\ Yoshiyasu Sato$^3$, Satoru Matsuishi$^4$, and Hideo Hosono$^5$}
\inst{
$^1$ Department of Engineering Science, 
The University of Electro-Communication, 
Chofu, Tokyo 182-8585, Japan\ \\
$^2$ Department of Physics,
Osaka University, 1-1 Machikaneyama, 
Toyonaka, Osaka 560-0043, Japan\ \\
$^3$ Materials and Structures Laboratory, 
Tokyo Institute of Technology, 
4259 Nagatsuta-cho, Midori-ku, Yokohama 226-8503, Japan\ \\
$^4$ Materials Research Center for Element Strategy, 
Tokyo Institute of Technology, 
4259 Nagatsuta-cho, Midori-ku, Yokohama 226-8503, Japan\ \\
$^5$ Frontier Research Center, Tokyo Institute of Technology, 
4259 Nagatsuta-cho, Midori-ku, Yokohama 226-8503, Japan\ \\
}

\abst{
We theoretically study the spin fluctuation and superconductivity in 
La1111 and Sm1111 
iron-based superconductors for a wide range of electron doping.
When we take into account the band structure variation by electron doping, 
the hole Fermi surface originating from the $d_{X^2-Y^2}$ orbital turns 
out to be robust against electron doping, and this gives rise to large spin 
fluctuations and consequently $s\pm$ pairing even in the heavily doped regime.
The stable hole Fermi surface is larger for Sm1111 than for La1111, which can 
be considered as the origin of the apparent difference in the phase diagram.
}

\kword{iron pnictide superconductors, Fermi surface, spin fluctuation, 
first-principles band calculation}

\begin{document}
\maketitle

The pairing mechanism of the iron-based superconductors 
has been of great interest ever since its discovery\cite{Kamihara2008}.
The possibility of spin-fluctuation-mediated pairing in the iron-based superconductors 
has been proposed from the very beginning of the study 
\cite{Mazin2008,KurokiPRL,KurokiBook}.  
The key point here is the presence of the disconnected electron 
and hole Fermi surfaces, which gives rise to spin fluctuations around the wave vector 
that connects the Fermi surfaces.The resulting pairing state 
is the so-called $s\pm$ state in which the sign of the gap function changes 
between electron and hole Fermi surfaces.

Recent observation of superconductivity in ``1111'' compounds 
{\it Ln}FeAsO$_{1-x}$H$_x$ ({\it Ln}=La,Sm, etc.) for a wide range of $x$ 
\cite{Iimura,Hanna} has raised an interesting issue regarding the possibility 
of spin-fluctuation-mediated pairing. There, the superconducting state persists 
up to a large $x$ of $\sim 0.4$, for which hole Fermi surfaces are expected 
to be completely wiped out in a rigid band picture. In fact, theoretical studies 
based on the
doping independent five orbital Hamiltonian show that the spin-fluctuation-mediated superconductivity
is lost for about 20 percent ($x$=0.2) electron doping\cite{Ikeda1,Ikeda2,KurokiPRB},
so that the observation of superconductivity in the heavily doped {\it Ln}FeAsO$_{1-x}$H$_x$
has lead to a proposal of the pairing mechanism based on local orbital fluctuations\cite{Ono}, 
in which the Fermi surface nesting does not play an important role.
Therefore, whether the spin fluctuation can be responsible for 
pairing even in such a heavily doped regime is certainly an intriguing problem.

To address this issue, in the present paper we study the 
spin fluctuations and superconductivity of 
La1111 and Sm1111 by applying random phase approximation (RPA) to 
a five orbital model that takes into account the effect of the 
variation of the band structure due to electron doping. 
It is found that the hole Fermi surface around the wave vector $(\pi,\pi)$ 
that originates from the $d_{X^2-Y^2}$ orbital is robust against electron 
doping, and this induces large spin fluctuations and $s\pm$ pairing even 
in the heavily electron doped regime. The stable hole Fermi surface 
is larger for Sm1111 than for La1111, and we propose that this is the 
origin of the difference in the experimentally observed 
$T_c$ vs. $x$ phase diagram, in which the latter material 
has a two-dome feature, while the former has a single-dome shape
\cite{Iimura,Hanna}.

We start with the first principles band calculation for the two materials.
The band calculation is performed 
using the VASP package\cite{VASP} for 
La1111 and Sm1111 compounds. Since the rigid band 
picture is expected to be unreliable especially in the heavily doped 
regime, we take into account the effect of the 
band structure variation due to doping by (i) 
adopting the lattice structure parameters determined experimentally for 
each doping rate $x$, and (ii) adopting  
 virtual crystal approximation in which 
the oxygen and the fluorine pseudopotentials are mixed, assuming  
that the hydrogen doping has an affect of increasing the 
average valence of the oxygen site by $+x$\cite{Iimura}.
We construct a five orbital tight-binding 
Hamiltonian in the unfolded Brillouin zone\cite{KurokiPRL}  
exploiting the maximally localized Wannier functions\cite{MaxLoc}. 
The five Wannier orbitals have five different symmetries ($d_{XY}$, $d_{YZ}$,
$d_{ZX}$, $d_{3Z^2-R^2}$ and $d_{X^2-Y^2}$), 
where $X,Y$ refer to the direction 
rotated by 45 degrees from the Fe-Fe direction $x,y$. 

\begin{figure}
\includegraphics[width=8.5cm]{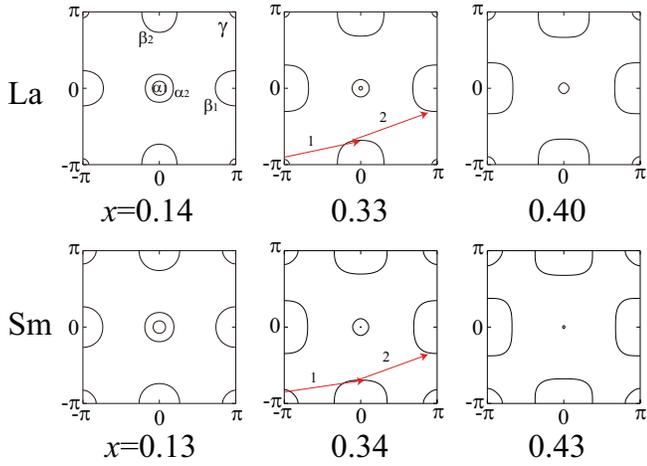}
\caption{(Color online) The Fermi surface for various doping rates for La1111 (top) and 
Sm1111 (bottom). The arrows 1 and 2 indicate the interactions between 
the Fermi surfaces which give rise to spin fluctuations and 
superconductivity.
}
\label{fig1}
\end{figure}

\begin{figure}
\includegraphics[width=8.0cm]{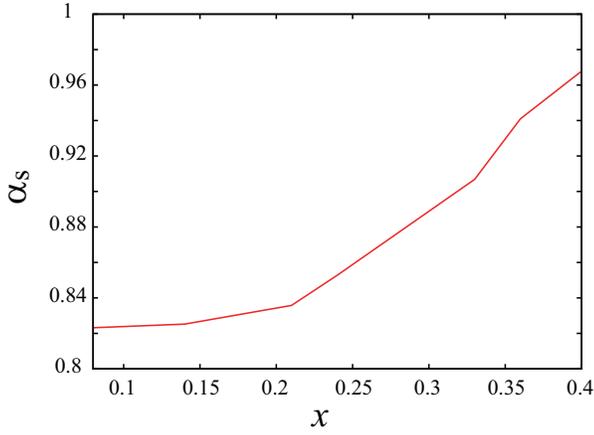}
\caption{(Color online) The Stoner factor for La1111 plotted against the doping rate. }
\label{fig2}
\end{figure}

We construct a many body Hamiltonian by taking into account the 
multi-orbital electron-electron interactions, and apply 
random phase approximation (RPA)\cite{Yada,Takimoto}. 
Using the Green's function $G_{lm}(k)$ $(k \equiv (\Vec{k},i\omega_n))$,
which is a $5\times 5$ matrix in the orbital representation, 
the irreducible susceptibility matrix is given as 
\begin{equation} 
\chi^0_{l_1,l_2,l_3,l_4}(q) =\sum_k G_{l_1l_3}(k+q)G_{l_4l_2}(k), 
\end{equation}
and the spin and the charge (orbital) susceptibility matrices are obtained as 
\begin{equation}
\hat{\chi}_s(q)=\frac{\hat{\chi}^0(q)}{1-\hat{S}\hat{\chi}^0(q)} ,
\end{equation}
\begin{equation}
\hat{\chi}_c(q)=\frac{\hat{\chi}^0(q)}{1+\hat{C}\hat{\chi}^0(q)} ,
\end{equation}
where $\hat{S}$ and $\hat{C}$ are the corresponding 
interacting vertex matrices. 
These matrices have $l_1$ to $l_4$ $(l_i = 1,...,5)$ as orbital indices. 
We calculate the Stoner factor defined as the largest 
eigenvalue of the matrix  $\hat{S}\hat{\chi}^0(q)$.
This qualitatively measures the strength of the low energy spin fluctuations, 
and a tendency toward magnetism.

The Green's function and 
the effective singlet pairing interaction, 
\begin{equation}
\hat{V}^s(q)=\frac{3}{2}\hat{S}\hat{\chi}_s(q)\hat{S}
-\frac{1}{2}\hat{C}\hat{\chi}_c(q)\hat{C}+\frac{1}{2}(\hat{S}+\hat{C}),
\end{equation}
are plugged into the linearized Eliashberg equation for superconductivity, 
\begin{eqnarray}
\lambda \phi_{l_1l_4}(k)&=&-\frac{T}{N}\sum_q
\sum_{l_2l_3l_5l_6}V_{l_1 l_2 l_3 l_4}(q)\nonumber\\
&\times& G_{l_2l_5}(k-q)\phi_{l_5l_6}(k-q)
G_{l_3l_6}(q-k).
\end{eqnarray}
The $5\times 5$ matrix gap function $\phi_{lm}$ 
in the orbital representation 
along with the associated eigenvalue $\lambda$ is obtained by 
solving this equation. Since $\lambda(T)=1$ signals $T=T_c$, 
the eigenvalue calculated at a fixed temperature is a qualitative measure for 
$T_c$. 
We present the gap function transformed into the band representation 
with a unitary transformation. 
The RPA calculation is performed for $64 \times 64 \times 4$ sites and 
2048 Matsubara frequencies, and the temperature is $T=0.02$eV. 
The electron-electron interactions are taken from ref. \citen{Miyake}, but 
we multiply all of them by a factor of $f=0.42$ since RPA overestimates the 
tendency towards magnetism.

\begin{figure*}[t]
\includegraphics[width=15cm]{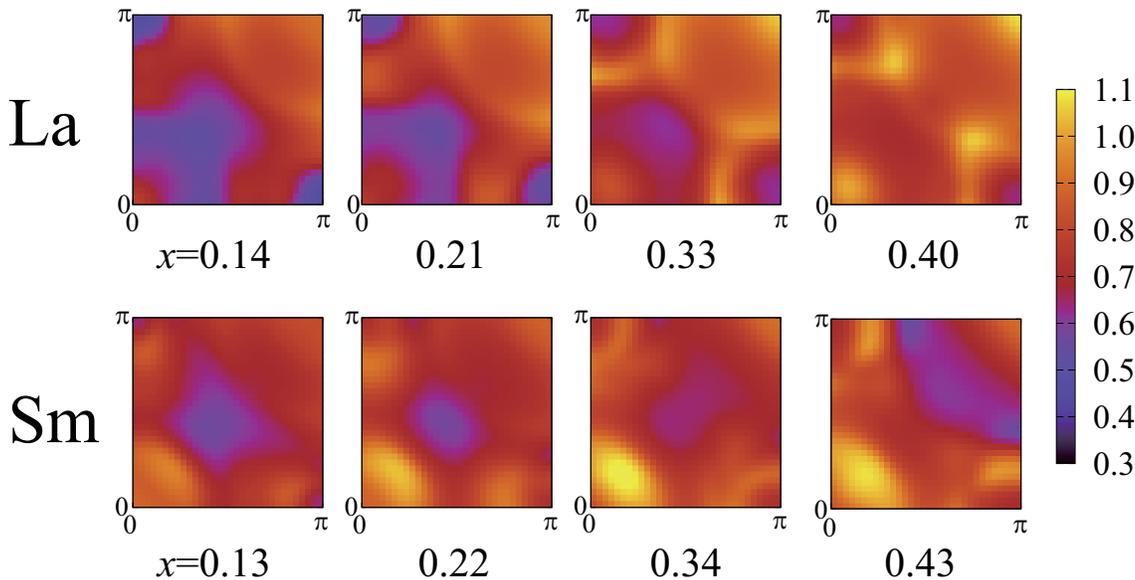}
\caption{(Color online) The ratio of the intra-orbital spin susceptibilities  
$\chi_{X^2-Y^2}^s/\chi_{XZ/YZ}^s$ for various doping rate for 
La1111(top) and Sm1111(bottom).}
\label{fig3}
\end{figure*}

We first show in Fig.\ref{fig1} the Fermi surface for various values of the 
doping rate $x$. The hole Fermi surfaces ($\alpha_1$, $\alpha_2$) 
around the wave vector $(0,0)$ 
originating from the $d_{XZ}/d_{YZ}$ orbitals 
shrink, while the electron Fermi surfaces ($\beta_1$, $\beta_2$) 
around $(0,\pi)/(\pi,0)$ become  
larger with doping, as expected. What is interesting is that the volume of the 
hole Fermi surface ($\gamma$) around $(\pi,\pi)$, which originates from the 
$d_{X^2-Y^2}$ orbital,  barely changes with doping 
for both materials. 
This is due to the band structure variation with doping, 
whose origin can be two folded. One is the lattice structure variation,
and the other is the change in the electric charge of the FeAs and {\it Ln}O 
layers. Both effects are taken into account in the present band calculation, 
and it turns out that the latter effect actually dominates. 
Namely, the energy level of $d_{XZ}/d_{YZ}$ orbitals (that 
are oriented toward the {\it Ln}O layers) is pushed down 
relative to $d_{X^2-Y^2}$ owing to the increase of the positive charge 
in the {\it Ln}O layers.
Comparing the two materials, 
the $\gamma$ Fermi surface is larger for Sm1111 due to higher pnictogen 
position\cite{KurokiPRB}.

\begin{figure}
\includegraphics[width=8.0cm]{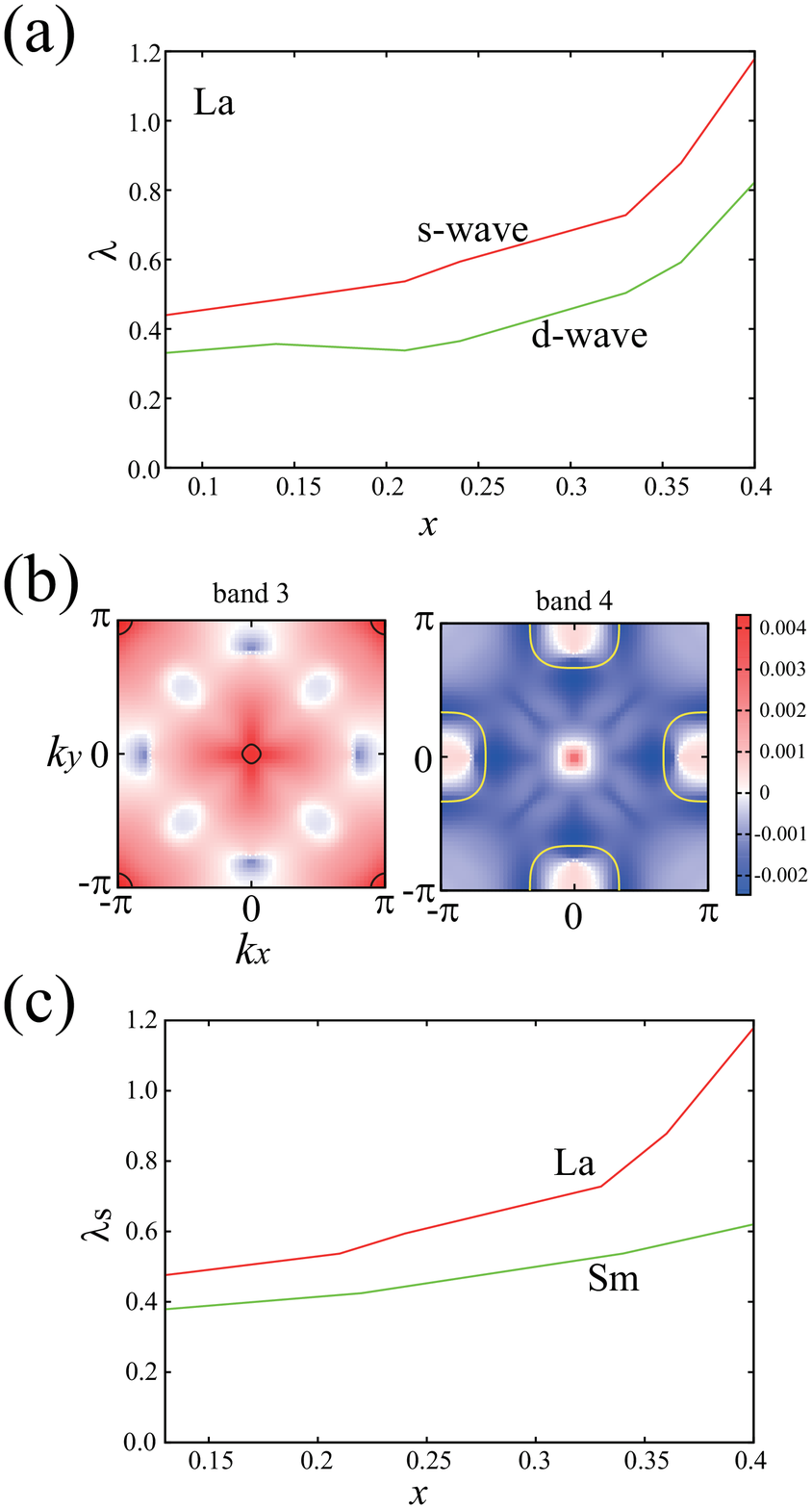}
\caption{(Color online) (a) The eigenvalue of the Eliashberg equation as 
a function of the doping for La1111. (b) The gap function for the $s$-wave 
pairing for band 3 (with the hole Fermi surfaces) and band 4 (the 
electron Fermi surfaces). The solid lines represent the Fermi surfaces. 
(c) The eigenvalue for the $s$-wave pairing against the doping rate for La and Sm. }
\label{fig4}
\end{figure}

In Fig.\ref{fig2}, we show the Stoner factor as a function of $x$. 
It can be seen that the Stoner factor actually increases with doping for 
a fixed electron-electron interaction, as opposed to a naive expectation 
based on ``rigid band + Fermi surface nesting'' picture.
To see this in more detail, we show in Fig.\ref{fig3} the ratio of the 
``intra-orbital spin susceptibility'' between $d_{X^2-Y^2}$ and $d_{XZ}/d_{YZ}$ 
orbitals\cite{KurokiPRB}. Here, the intra-orbital spin susceptibility is 
defined as the diagonal element of the spin susceptibility matrix that has the 
same orbital indices. 
In La1111, the $d_{XZ}/d_{YZ}$ susceptibility dominates 
for small doping, but as the doping rate is increased, $d_{X^2-Y^2}$ tends 
to dominate, particularly around the wave vector 
$(\pi,\pi/3)/(\pi/3,\pi)$\cite{comment}.
The contribution to this wave vector comes from two interactions;
one from the $\gamma$-$\beta$ (electron-hole, arrow 1 in 
Fig.\ref{fig1}) interaction, 
and another from $\beta_1$-$\beta_2$ (electron-electron, arrow 2)
interaction\cite{KurokiPRB,Graser,DHLee,Thomale}. These two contributions,
both coming mainly from the intra-orbital repulsion within the 
$d_{X^2-Y^2}$ orbitals, cooperate to 
give a large enhancement in the incommensurate spin fluctuations. 
In fact, the enhancement of the spin fluctuation around this wave vector 
is consistent with a recent experimental study\cite{Iimura2}.
The cooperation occurs because the two contributions accidentally 
have similar incommensurate wave vectors especially for large $x(\sim 0.3-0.4)$.
As for the material dependence, 
the difference between La1111 and Sm1111 can be seen by comparing the 
upper and lower panels in Fig\ref{fig3}.
While a crossover from $d_{XZ}/d_{YZ}$ to $d_{X^2-Y^2}$ dominating 
regimes is seen in La, the $d_{X^2-Y^2}$ contribution is large through the 
entire doping regime for Sm1111. 
This is due to the presence of the larger $\gamma$ 
Fermi surface in the latter material.

The two origins for the spin fluctuations, i.e., the $\gamma$-$\beta$ 
and $\beta_1$-$\beta_2$  interactions, 
cooperate toward antiferromagnetic instability, but 
they have competing effects regarding the occurrence of superconductivity. 
Namely, as discussed in previous 
studies\cite{KurokiPRL,KurokiPRB,Graser,DHLee,Thomale}, 
the $\gamma$-$\beta$ interaction leads to $s\pm$ pairing, while 
the $\beta_1$-$\beta_2$ interaction favors $d$-wave pairing because this 
interaction acts to change the gap sign between $\beta_1$ and $\beta_2$. 
To see which one of the interactions dominate in the pairing interaction, 
we calculate the eigenvalue of the Eliashberg equation for $s$ and $d$-wave pairings. 
As shown in Fig.\ref{fig4}(a), $s$-wave always dominates over $d$-wave, 
and the gap function has the fully gapped $s\pm$ form 
as shown in Fig.\ref{fig4}(b), where the gap changes sign between the electron 
and hole Fermi surfaces. 

The present result suggests that the $d_{X^2-Y^2}$ hole Fermi surface 
plays an important role in the occurrence of superconductivity even 
in the heavily doped regime. One should note, however, that the nesting 
(in the conventional sense of the term) between the electron and the 
hole Fermi surfaces is not good, since the volume is very different. 
Nevertheless, the pairing is strongly enhanced. This is because the 
portions of the bands away from the Fermi level contribute to the pairing 
interaction through finite energy virtual processes. Therefore, 
one should be careful about using the word ``Fermi surface nesting'' as the 
origin of the pairing interaction, at least in the naive sense of the term.

The material dependence seen in the intra-orbital contribution to the 
spin susceptibility is also reflected to superconductivity.
In Fig.\ref{fig4}(c), we plot the eigenvalue of the Eliashberg equation against 
the doping rate for the two materials.
We see that it is strongly enhanced in the heavily doped regime for La1111, 
while the variation is relatively weak for Sm1111
This is because the contribution weight to the spin fluctuations from the $d_{X^2-Y^2}$ 
orbital is less dependent on the doping rate in Sm compared to that in La. 
The weak doping dependence of $\lambda$  in Sm1111 is 
qualitatively consistent with the small variance of $T_c$ for a wide range of $x$ 
observed in Sm1111\cite{Hanna}.

Here, we should comment on some discrepancies with the experimental observations. 
In both cases, $\lambda$ monotonically increases with doping, 
and $\lambda$ for La is larger than Sm. 
In the actual experiments, $T_c$ has two (single) dome shape dependence 
against doping in La (Sm), and the maximum $T_c$ is higher for Sm. 
Our present understanding is that this discrepancy comes from overestimating 
the effect of the density of states (DOS) variance in the RPA calculation. 
Namely, the band width decreases with doping and is also smaller in La than in Sm, 
which  is directly reflected in the strength of the spin fluctuations and thus $\lambda$. 
In reality, the increase in DOS enhances the self energy correction (neglected in RPA), 
which suppresses the spin fluctuation and $\lambda$. 
Thus our expectation is that such a DOS variance overestimation effect is reduced in calculation 
that considers self energy corrections, which is now underway.
Although a complete understanding of the shape of the phase diagram lies beyond the present 
theoretical approach, smaller eigenvalues obtained for small $x$ in La1111 suggests a fragile 
nature of the superconductivity in this regime.

To summarize, we have analyzed the spin fluctuation and superconductivity 
of the 1111 iron pnictides in a wide range of electron doping rate  
by taking into account the 
band structure variation within the virtual crystal approximation. 
The $d_{X^2-Y^2}$ hole Fermi surface around 
$(\pi,\pi)$ is found to be robust for the entire doping regime, and plays an 
important role on both the spin fluctuations and superconductivity.
In La1111, the $d_{XZ}/d_{YZ}$ orbitals have relatively large contribution in the 
small $x$ regime, while in Sm1111, the $d_{X^2-Y^2}$ dominates 
over the entire regime. This difference in the $d_{X^2-Y^2}$ contribution 
is plausibly the origin of the difference in the phase diagram between the two materials.
After completion of this study, a paper that studies a similar problem for 
LaFeAsO$_{1-x}$H$_x$ has been posted on the arXiv\cite{Yamakawa}.

The numerical calculations were performed at the Supercomputer Center, 
ISSP, University of Tokyo. This study has been supported by 
Grant-in-Aid for Scientific Research No.24340079 from JSPS. 
K.S. acknowledges support from JSPS. The part of the research 
at Tokyo Institute of Technology was supported by the JSPS FIRST Program.

\bibliography{perovskite}

\begin{thebibliography}{99}
\bibitem{comment} There is also an ``enhancement'' of the 
$\chi^s_{X^2-Y^2}$ vs. $\chi^s_{XZ/YZ}$ ratio around the 
wave vector (0,0), especially for Sm1111. Note, however, that this does 
not imply enhanced ferromagnetic fluctuations because it is the 
{\it ratio} that is enhanced, not the bare values. The bare values around 
(0,0) is small compared to those around $(\pi,0)\sim (\pi,\pi/3)$.
\end{thebibliography}

\end{document}